\documentclass[onecolumn,showpacs]{revtex4}
\usepackage{graphicx}
\usepackage{dcolumn}
\usepackage{bm}
\usepackage{amsmath}
\usepackage{amssymb}
\usepackage{epstopdf}
\usepackage{hyperref}

\begin{document}
{\setlength{\oddsidemargin}{1.2in}
\setlength{\evensidemargin}{1.2in} } \baselineskip 0.55cm
\begin{center}
{\LARGE {\bf Anisotropic Universe in $f(Q)$ gravity with Hybrid expansion}}
\end{center}
\date{\today}
\begin{center}
L. Anjana Devi, S. Surendra Singh, Leishingam Kumrah and Md Khurshid Alam \\
   Department of Mathematics, National Institute of Technology Manipur,\\ Imphal-795004,India\\
   Email:{ anjnlam@gmail.com, ssuren.mu@gmail.com , lkumrah@gmail.com and alamkhurshid96@gmail.com}\\
 \end{center}

\textbf{Abstract}: Despite having a reasonably successful account of accelerated cosmology, understanding the early evolution of Universe has always been difficult for mankind. Our promising strategy is based on a novel class of symmetric teleparallel theories of gravity called $f(Q)$,  in which the gravitational interaction is caused by the non-metricity scalar $Q$, which may help to solve some problems. We consider the locally rotationally symmetric (LRS) Bianchi type-I spacetime cosmological models and derive the motion of equations to study the early evolution of the cosmos. By assuming the Hybrid Expansion Law (HEL) for the average scale factor, we are able to determine the solutions to the field equations of Bianchi type-I spacetime. We discuss the energy density profile, the equation of state, and the skewness parameter and conclude that our models preserve anisotropic spatial geometry during the early stages of the Universe with the possibility of an anisotropic fluid present. However, as time goes on, even in the presence of an anisotropic fluid, the Universe may move towards isotropy due to inflation while the anisotropy of the fluid dims away at the same time. It is seen from the squared speed of sound that Universe shows phantom nature at the beginning then approaches to dark energy at present epoch. We analyze both geometrical and physical behaviors of the derived model.


\section{Introduction}\label{sec1}

The current acceleration epoch of the Universe is accelerated, according to observations of high-redshift supernovae and cosmic microwave background fluctuations (CMBR) \cite{1,2,3,4}. This late time acceleration is due to an unknown fluid called dark energy (DE). Several concepts have been considered as a candidate to explain the true nature of DE. The first is the cosmological constant that encounters problems such as impossibly small value required by general relativity (GR) theory. Also, the cosmological constant is many orders of magnitude smaller than estimated in modern theories of elementary particle physics in which predictions is generally more that 50 orders of magnitude than the actual calculation by GR \cite{5}. This strange DE that is responsible for the cosmological acceleration and is estimated as 68\% of the total energy density of the Universe whcih may be in need of us to reconsider the theory of gravity on cosmological scales. The DE can be examined using an effective tool namely the equation of state (EoS) parameter of the form $w=\frac{p}{\rho}$, which is the ratio of the cosmic pressure $p$ to the cosmic energy density $\rho$. Each DE model has a different EoS parameter value, for instance, in the case of the cosmological constant its value is $w=-1$, also for the quintessence model $w$ is bounded as $-1<w<-0.33$, and lastly $w<-1$ for the phantom DE model.\\
Modified gravity theories provide the interesting theoretical concepts to the notion of cosmological constant problem and describe the late-time acceleration of the Universe. Many DE models which are started from the simplest modified gravity $\frac{1}{R}$ theory \cite{6,7}. In general, modified gravity theories appear to be quite interesting since it provides instinctive solutions to a number of key problems concerning DE. The another theory to GR is teleparallel gravity by which gravitational interaction is described by the torsion scale $T$ \cite{8,9,10} in a space-time with zero curvature. This theory is entitled teleparallel equivalent to general relativity $(TEGR)$ and formulated by tetrad fields which is different from the Levi-Civita interrelation in GR. The superiority of working with $f(T)$ models is the order of the field equations, this permits simplifying the dynamics and finding simple exact solutions. The covariant derivative of the metric tensor does not vanish in the alternative theory of symmetric teleparallel $f(Q)$ gravity i.e. $Q_{\gamma\mu{v}}=\Delta_{\gamma}g_{\mu{v}}$. This theory is called symmetric teleparallel equivalent to general relativity $(STEGR)$ \cite{11,12}. This new modified $f(Q)$ gravity where $Q$ is the non-metricity scalar attracted the interest of several researchers \cite{13,14,15,16,17,18}. Further, this theory is established on the generalization of Riemannian geometry described by Weyl's geometry \cite{19}. Generally, the gravitational interaction is categorized through three types of geometries: the curvature of space-time, torsion and non-metricity. Accordingly, in the latest decades researchers have been interested to modified gravity theories because they reflect the current phenomena of the Universe. Therefore, gravitational interactions have been considered using several forms of geometries \cite{20,21,22}.\\

  When the inflationary phase was successfully established, observations showed that the Universe was homogeneous and isotopic \cite{23}. However, inconsistencies in the CMBR lead to conclude that an anisotropic phase in the early Universe which construct it, is not exactly uniform \cite{24}. Therefore, it is essential to take into consideration the anisotropic and inhomogeneous characteristics of the Universe while developing the cosmological models. In order to achieve this, Bianchi-type models provide a reasonable description of the anisotropic background and examine the cosmic evolution in the early Universe. In fact, there are nine different types of Bianchi models that described in the literature. Here, we consider the anisotropic locally rotationally symmetric (LRS) Bianchi type-I model which is presumed to be a more extended cosmological metric than Friedmann-Lemaitre-Robertson-Walker (FLRW) metrics \cite{25}. The Bianchi type-I model is used to test the potential consequences of anisotropy in the early Universe \cite{26}. Lately, anisotropic fluid in Bianchi type-I space-time has been used to construct cosmological models. Additionally, some  Bianchi type-I solutions have also been studied in $f(Q)$ modified gravity \cite{27,28}. The Bianchi type-I model generally presents good consistency with the most basic mathematical form, considering the nature of this model. Bianchi type-I theory was investigated in the setting of a viscous fluid to discuss the behavior of the early Universe near the singularity \cite{29}.\\

  Due to the dependability of the deceleration parameter, there is also considerable interest in studying the Universe with power-law and exponential law expansion to explain epoch-based development of the Universe. This study requires the transitional phase of the Universe from the epoch of slowdown expansion to the epoch of acceleration. However, in the models with Hybrid Expansion Law (HEL), we obtain the transition from deceleration to acceleration in the evolution of the Universe. In Lyra's manifold and scale covariant theory of gravity, Ibotombi et al. \cite{30,31} examined the power law and exponential law from bulk viscous cosmological models. According to Surendra et al. \cite{32}, who explored power-law inflation in Lyra's manifold with an anisotropic fluid, the Universe is anisotropic at first and becomes isotropic afterwards. The HEL, which is the outcome of power-law and exponential-law functions, is a simple depiction of the Universe's growth history within the context of the Brans-Dicke hypothesis \cite{33}. The hybrid model is  appropriate for the phase change from deceleration to acceleration, according to recent talks of another approach to the dark energy interacting with the hybrid model in $f(R,T)$ gravity \cite{34}. It is established that the values of matter-energy density $(\Omega_{m})$ and dark-energy density $(\Omega_{\Lambda})$ are consistent with five years of WMAP measurements by demonstrating the dynamical nature of anisotropic dark energy in $f(R,T)$ gravity.\\

The preceding work served as an inspiration for the current research, which focuses on studying LRS Bianchi type-I space time with interacting anisotropic dark energy in $f(Q)$ gravity. We assumed a scale factor that took the form of a hybrid expansion law to solve the generalized field equations. The structure of the current paper is as follows: in section \ref{sec2}, we obtain the equations for $f(Q)$ gravity and its field equations. In section \ref{sec3}, we go over the model using a HEL. We study stability and energy conditions in section \ref{sec4}. Section \ref{sec5} analyzes $Om(z)$ diagnostic of the model. Finally, in section \ref{sec6}, we examine our findings and provide a conclusion.

\section{Metric and Field Equations}\label{sec2}

  We examine a special type of Bianchi Universe, the locally rotationally symmetric (LRS) Bianchi type-I model to indicate the anisotropic state of the Universe, given by the metric in the Cartesian coordinates
\begin{equation}\label{1}
 ds^2=dt^2-A^2dx^2-B^2(dy^2+dz^2).
\end{equation}
Here $A$ and $B$ are the scale factors, which only depend on cosmic time $t$. The action for symmetric teleparallel gravity is represented in the following form
\begin{equation}\label{2}
 S= \int \frac{1}{2k}f(Q)\sqrt{-g} dx^4 + L_{m}\sqrt{-g} dx^4,
 \end{equation}
where $k=8\pi{G}=1$, $f(Q)$ can be expressed as the arbitrary function of non-metricity scalar $Q$, $g$ is the determinant of the metric tensor $g_{\mu {v}}$, and $L_{m}$ is the Lagrangian density. Now, the non-metricity tensor $Q_{\gamma\mu{v}}$ and its traces can be written as
\begin{equation}\label{3}
Q_{\gamma\mu{v}}=\Delta_{\gamma}g_{\mu{v}}
\end{equation}
\begin{equation}\label{4}
Q_{\gamma}=Q_{\gamma\mu}^{\mu}~~,~~~~\overset{\sim}Q_{\gamma}=Q^{\mu}_{\gamma\mu}
\end{equation}
In addition, the superpotential tensor (non-metricity conjugate) can be expressed as
\begin{equation}\label{5}
4P^{\gamma}_{\mu{v}}=-Q^{\gamma}_{\mu{v}}+2Q_{\mu{v}}^{\gamma}-Q^{\gamma}g_{\mu{v}}-\overset{\sim}Q^{\gamma}g_{\mu{v}}-\delta^{\gamma}_{\mu}Q_{v}
\end{equation}
where the trace of the non-metricity tensor can be obtained as
\begin{equation}\label{6}
Q=-Q_{\gamma\mu{v}}P^{\gamma\mu{v}}
\end{equation}
Now, the matter energy-momentum tensor is defined as
\begin{equation}\label{7}
T_{\mu{v}}=-\frac{2}{\sqrt{-g}}\frac{\delta(\sqrt{-g}L_{m})}{\delta g^{\mu{v}}}
\end{equation}
By varying the modified Einstein-Hilbert action \eqref{2} with respect to the metric tensor $g_{\mu{v}}$, the gravitational field equations are obtained as

\begin{equation}\label{8}
\frac{2}{\sqrt{-g}}\Delta_{\gamma}(\sqrt{-g}f_{Q}P^{\gamma}_{\mu v})-\frac{1}{2}fg_{\mu v}+f_{Q}(P_{v\rho\sigma}Q_{\mu}^{\rho\sigma}-2P_{\rho\sigma\mu}Q^{\rho\sigma}_{v})=kT_{\mu v}.
\end{equation}
 where $f_{Q}=\frac{df}{dQ}$. The EoS parameter of the gravitational fluid should, in theory, also be generalized to reveal an anisotropic properties in order to give a more reasonable model. To study non-trivial isotropization in the expansion of the Universe, the anisotropic type space-time metric is taken into examination. The fluid also isotropizes with the isotropization of the Universe, showing an isotropic pressure and vanishing skewness parameter. The energy-momentum tensor for the anisotropic fluid is\\

\begin{equation}\label{9}
  T^i_j=diag\Big[\rho,-p_x,-p_y,-p_z\Big],
 \end{equation}
where $\rho$ stands for the fluid energy density, $p_{x},p_{y}$ and $p_{z}$ are pressures along  $x, y$ and $z$ coordinates respectively which are assumed to correspond to the directional equation of state parameters $w_{x},w_{y}$ and $w_{z}$. The deviation from isotropy is parametrized by assigning  $w_{x}=w_{y}=w_{z}=w$ and then introducing the deviations along $y$ and $z$ directions by the skewness parameter $\delta$, where $w$ and $\delta$ are functions of time.  Using the metric \eqref{1}, the components of $T_{ij}$ are given by the following:
\begin{equation}\label{10}
T_{00}=\rho, T_{11}=A^2p_{x}, T_{22}=B^2p_{y}, T_{33}=B^2p_{z}.
\end{equation}
  We derive the expansions of $\rho, p_{x},p_{y},p_{z},w, \delta$. Using Eqs. \eqref{1} and \eqref{8}, we find the following equations of motion:
\begin{equation}\label{11}
\rho= \frac{f}{2}+f_{Q}\Big[4\frac{\dot{A}}{A}\frac{\dot{B}}{B}+2\Big(\frac{\dot{B}}{B}\Big)^2\Big],
\end{equation}
\begin{equation}\label{12}
p_{x}=-\frac{f}{2}+f_{Q}\Big[-2\frac{\dot{A}}{A}\frac{\dot{B}}{B}-2\frac{\ddot{B}}{B}-2\Big(\frac{\dot{B}}{B}\Big)^2\Big]-2\frac{\dot{B}}{B}\dot{Q}f_{QQ},
\end{equation}
\begin{equation}\label{13}
p_{y}=-\frac{f}{2}+f_{Q}\Big[-3\frac{\dot{A}}{A}\frac{\dot{B}}{B}-\frac{\ddot{A}}{A}-\frac{\ddot{B}}{B}-\Big(\frac{\dot{B}}{B}\Big)^2\Big]-\Big(\frac{\dot{A}}{A}+\frac{\dot{B}}{B}\Big)\dot{Q}f_{QQ},
\end{equation}
\begin{equation}\label{14}
p_{z}=p_{y}.
\end{equation}
where the overhead dot (.) indicates the derivative with respect to cosmic time $t$. From the consideration of the anisotropic fluid
\begin{eqnarray}\label{15}
  T^{\mu}_{v}&=&diag\Big[\rho,-p_{x},-p_{y},-p_{z}\Big],\cr
              &=&diag\Big[1,-w_{x},-w_{y},-w_{z}\Big]\rho,\cr
              &=&diag\Big[1,-w,-(w+\delta),-(w+\delta)\Big]\rho,
\end{eqnarray}
The equation of motion can be expressed as

\begin{equation}\label{16}
\rho= \frac{f}{2}+f_{Q}\Big[4\frac{\dot{A}}{A}\frac{\dot{B}}{B}+2\Big(\frac{\dot{B}}{B}\Big)^2\Big],
\end{equation}
\begin{equation}\label{17}
w\rho=-\frac{f}{2}+f_{Q}\Big[-2\frac{\dot{A}}{A}\frac{\dot{B}}{B}-2\frac{\ddot{B}}{B}-2\Big(\frac{\dot{B}}{B}\Big)^2\Big]-2\frac{\dot{B}}{B}\dot{Q}f_{QQ},
\end{equation}
\begin{equation}\label{18}
(w+\delta)\rho=-\frac{f}{2}+f_{Q}\Big[-3\frac{\dot{A}}{A}\frac{\dot{B}}{B}-\frac{\ddot{A}}{A}-\frac{\ddot{B}}{B}-\Big(\frac{\dot{B}}{B}\Big)^2\Big]-\Big(\frac{\dot{A}}{A}+\frac{\dot{B}}{B}\Big)\dot{Q}f_{QQ}.
\end{equation}
    We have the directional Hubble parameters in the direction of the $x,y$ and $z$, correspondingly, are given by
    \begin{equation}\label{19}
    H_{x}=\frac{\dot{A}}{A}, H_{y}=\frac{\dot{B}}{B}, H_{z}=\frac{\dot{B}}{B},
    \end{equation}
 and the average Hubble parameter, which intimate the volumetric expansion rate of the Universe is given by
 \begin{equation}\label{20}
 H=\frac{1}{3}\frac{\dot{V}}{V}=\frac{1}{3}\Big[\frac{\dot{A}}{A}+2\frac{\dot{B}}{B}\Big],
 \end{equation}
 where the spatial volume as
 \begin{equation}\label{21}
 V=AB^2.
 \end{equation}

The mean anisotropy parameter $\Delta$, shear scalar $\sigma^2$ and the expansion scalar $\theta$ of the fluid are defined as follows
\begin{equation}\label{22}
\Delta=\frac{1}{3}{\sum\Big({\frac{H_{i}-H}{H}}}\Big)^2,
\end{equation}
\begin{equation}\label{23}
       {\sigma}^2={\frac{1}{2}}\Big[{\sum H_i^2}-3H^2\Big]
       ={\frac{3}{2}}{\Delta}H^2\,
       \end{equation}
\begin{equation}\label{24}
        \theta=\frac{\dot{A}}{A}+2\frac{\dot{B}}{B}.
        \end{equation}
In order to simplify the form of the field equations \eqref{11}-\eqref{14}and write them in terms of the non-metricity scalar $Q$, the average Hubble parameters $H$, and the directional Hubble parameters $H_{x}, H_{y}$. We note that $\frac{\delta}{\delta t}\Big(\frac{\dot{A}}{A}\Big)=\frac{\ddot{A}}{A}-\Big(\frac{\dot{A}}{A}\Big)^2$ and $Q=-2H_{y}^2-4H_{x}H_{y}$. In general relativity, we explore the condition $f(Q)=Q$ to examine the LRS Bianchi type-I Universe. In this case the equations of motion \eqref{16}-\eqref{18} reduce to
\begin{equation}\label{25}
\rho=-\frac{Q}{2},
\end{equation}
\begin{equation}\label{26}
w\rho=-\frac{Q}{2}-2\Big[\frac{\dot{A}}{A}\frac{\dot{B}}{B}+\Big(\frac{\dot{B}}{B}\Big)^2+\frac{\ddot{B}}{B}\Big],
\end{equation}
\begin{equation}\label{27}
(w+\delta)\rho=-\frac{Q}{2}-\Big[3\frac{\dot{A}}{A}\frac{\dot{B}}{B}+\Big(\frac{B}{B}\Big)^2+\frac{\ddot{A}}{A}+\frac{\ddot{B}}{B}\Big].
\end{equation}
Lastly, here we have three equations with five unknowns $A,B,\rho, w$ and $\delta$. The exact solutions of these equations are calculated in the next section.\\

\section{Solution of the field equations}\label{sec3}

To analyze the field equations of the $f(Q)$ gravity, we consider the scale factor in the form of HEL.
     \begin{equation}\label{28}
     a=a_{o} t^{n}e^{{\alpha}t}.
     \end{equation}

 In this place, here $a_{o}>0$, $n\geq0$ and $\alpha\geq0$ are constants. This is the generalized form of the scale factor which unifies power-law and exponential expansions of the Universe. With $\alpha=0$ and $n=0$, respectively, we can examine the power-law expansion and the exponential growth. Thus $n>0$ and $\alpha>0$ give a new cosmology derived from HEL. The Universe shows initial singularity since $a{\sim}0$ at $t{\sim}0$, and evolves with infinite expansion since  $a{\sim}{\infty}$ at $t\sim{\infty}$. Using \eqref{28}, we secure the expression of deceleration parameter $q$ as

 \begin{equation}\label{29}
     q=\frac{n}{({\alpha}t+n)^2}-1.
  \end{equation}
We observed that the deceleration parameter is a function of time $t$, if $\alpha$ and $n$ are positive, and a change from the deceleration to the acceleration phase occurs at $t=\frac{\Big(\sqrt{n}-n\Big)}{\alpha}$ , which limits $n$ to the range  $0<n<2$. The equation \eqref{21}, becomes\\
      \begin{equation}\label{30}
      V=a^3=AB^2=a_{o}^{3}t^{3n}e^{3{\alpha}t}.
     \end{equation}
With the help of preceding expression and assuming the condition that the expansion scalar be proportional to the shear scalar i.e $A=B^m$, to find a solution in the form of Hybrid expansion law, which gives
     \begin{equation}\label{31}
     A=a_{o}^{\frac{3n}{n+2}}t^{\frac{3n^2}{n+2}} e^{\frac{3n\alpha t}{n+2}},
         \end{equation} and
         \begin{equation}\label{32}
 B=a_{o}^{\frac{3}{n+2}}t^{\frac{3n}{n+2}} e^{\frac{3\alpha t}{n+2}}.
\end{equation}
 Using the above expression, the directional Hubble parameters in this model are found to be in the following form
   \begin{equation}\label{33}
        \frac{\dot{A}}{A}= H_{x}=\frac{3n}{n+2}\Big(\alpha+\frac{n}{t}\Big),
         \end{equation}
         ~~~~~~and~~~~~~~
         \begin{equation}\label{34}
         \frac{\dot{B}}{B}= H_{y}=H_{z}=\frac{3}{n+2}\Big(\alpha+\frac{n}{t}\Big).
         \end{equation}
 We observed that directional Hubble parameters are very large at the initial stage of expansion and very small at later stage of evolution. This explains how the Universe is initially anisotropic and gradually becomes isotropic as it continues to age. The average Hubble's parameter ($H$) in this model is obtained as
        \begin{equation}\label{35}
        H= \alpha+\frac{n}{t}.
         \end{equation}
For $\alpha>0$ and $n>0$, the Hubble parameter $H$ is always positive. The anisotropy parameter $\Delta$, shear scalar $\sigma^2$ and the scalar expansion $\theta$ of our model are obtained as
\begin{equation}\label{36}
    \Delta=\frac{1}{3}{\sum\Big({\frac{H_{i}-H}{H}}}\Big)^2
    =\frac{2(n-1)}{(n+2)^2}.
\end{equation}
\begin{equation}\label{37}
       {\sigma}^2={\frac{1}{2}}\Big[{\sum H_i^2}-3H^2\Big]
       ={\frac{3}{2}}{\Delta}H^2
       =\frac{3(n-1)(\alpha+\frac{n}{t})^2}{(n+2)^2}.
       \end{equation}
~~~~~~and~~~~~
        \begin{equation}\label{38}
        \theta=3H=3\Big(\alpha+\frac{n}{t}\Big)=\frac{\dot{A}}{A}+2{\frac{\dot{B}}{B}}.
       \end{equation}
Here, we can see that the Hubble's parameter, shear scalar and scalar expansion diverge at $t=0$. It is also feasible to look at the isotropic condition $\frac{\sigma^2}{\theta^2}$, as it takes a constant value from the initial time to the late time. Therefore, our model suggest that, it does not come close to the isotropy throughout the expansion of the Universe, and this is confirmed by Eq. \eqref{36}, where we observed that the anisotropy parameter is constant for our model. Using \eqref{33} and \eqref{34} in \eqref{25}, we obtained the expressions of energy density as
        \begin{equation}\label{39}
\rho=\frac{9(1+2n)}{(n+2)^2}\Big(\alpha+\frac{n}{t}\Big)^2.
\end{equation}
Now, using \eqref{33}, \eqref{34} and \eqref{39} in \eqref{26} and \eqref{27}, we obtained the expression of the EoS parameter and skewness parameter in our model as
      \begin{equation}\label{40}
 w=-1+\frac{2n{t^2 \Big(\alpha+\frac{n}{t}\Big)^2+4t^2 \Big(\alpha+\frac{n}{t}\Big)^2}-6n(n+2)}{t^2(1+2n)\Big(\alpha+\frac{n}{t}\Big)^2},
     \end{equation}
     ~~~~~~and~~~~~~
     \begin{equation}\label{41}
  \delta= \frac{\Big[2-n-\frac{n(n+2)}{3t^2 \Big(\alpha +\frac{n}{t}\Big)^2}-n^2+\frac{n^2(n+2)}{3t^2\Big(\alpha+\frac{n}{t}\Big)^2}\Big]}{(1+2n)}.
  \end{equation}
 Using the correlation of equation of state, $(p=\omega \rho)$, we have the expression of pressure as
 \begin{equation}\label{42}
  p=9\Big(\alpha + \frac{n}{t}\Big)^2\Big[\frac{1+2n}{(n+2)^2}-2n-2+\frac{6n(n+2)}{t^2 \Big(\alpha +\frac{n}{t}\Big)^2}\Big].
  \end{equation}
We are taking the two potential values of $\alpha$ and $n$ in this case. Finding appropriate values for these characteristics is all that is necessary to expand the physically sound cosmological theories. As a result, we use the range of $0<\alpha<1$ and $0<n<2$ to make the figures interesting. According to their different behaviours with the model's evolution, the deceleration parameter and Hubble expansion parameter are shown in Fig.1 and Fig.2 respectively. We noticed that the Hubble parameter is dropping with time before stabilizing at a late point. Furthermore, the deceleration parameter behaves positively at first and subsequently more negatively  at the late time. This illustrates how the Universe changed from an acceleration phase. Fig.3 shows that the energy density is always positive as the Universe expands, which is consistent with the observations made at the time. As seen in Fig. 4, the EoS parameter is always negative i.e $0>w >-1$.

\section{ Energy conditions and stability behaviour of the proposed model}\label{sec4}

{\bf IV.I Energy conditions of the model:}
 Setting $a(t)=1/(1+z)$, where $z$ is the redshift, results in the expression for the relationship between time and redshift as
   \begin{equation}\label{43}
   t=\frac{n}{\alpha}W\Big[\frac{\alpha}{n}\Big(\frac{1}{a_{0}(1+z)}\Big)^{1/n}\Big],
   \end{equation}
Here, $W$ stands for the Lambert $W$ function. The parameters of the developed model can be represented in terms of redshift using relation \eqref{43}. A relationship like this is helpful when using observational data to test the model. Energy conditions in general relativity are divided into four categories: null energy condition (NEC), weak energy condition (WEC), strong energy condition (SEC), and dominant energy condition (DEC), each of which is represented as
 \begin{equation}\label{44}
NEC\Leftrightarrow{\rho}+p\geq0,
\end{equation}

\begin{equation}\label{45}
WEC{\Leftrightarrow} NEC ~~~~~and~~~~~  \rho\geq0,
\end{equation}

\begin{equation}\label{46}
SEC\Leftrightarrow \rho+3p \geq0,
\end{equation}

\begin{equation}\label{47}
DEC\Leftrightarrow \rho-p \geq0.
\end{equation}
 We noticed from Figures 3, 7, 8 and 9 that $\rho>0$, $\rho-p>0$, $\rho+p>0$ and $\rho+3p<0$. The strong energy condition (SEC) is violated, but the null energy condition (NEC), weak energy condition (WEC), and dominant energy condition (DEC) are all satisfied. Many writers have proposed the same energy condition (Alvarenga et al., 2013) and ( Sharif et. al., 2013). Observations show that dark energy will dominate in the expanding Universe in the future.

{\bf IV.II  The squared sound speed:}
For the stability of our model, here, we analysis the adiabatic squared sound speed. It is one of the significant factors in cosmology. For any fluid, the squared sound speed is defined as
\begin{equation}\label{48}
v_{s}^2= \frac{\partial{p}}{\partial{\rho}},
\end{equation}

Hence, $v_{s}^2$ has three possibilities i.e $v_{s}^2<0$ or $v_{s}^2=0$ or $v_{s}^2>0$. Investigation of the sign of $v_{s}^2$ is important since it indicates the cosmological model's instability, which allows one to accept or reject their setup. If $v_{s}^2<0$, the energy density perturbation will increase beyond control and cause the cosmological models to be classically unstable. If $v_{s}^2>0$, this could provide a problem with the possibility of a causality. As a matter of fact, it is frequently evaluated as $v_{s}\leq1$ and bound on $v_{s}^2$ is $0\leq v_{s}^2 \leq 1$. Additionally, the complementary bound $v_{s}>1$ is employed as a requirement for disproving the theories. The details regarding $v_{s}^2$ can be found in \cite{37}. In our present model, $v_{s}^2$ is obtained as
\begin{eqnarray}\label{49}
v_{s}^2&=&-\frac{1}{2(1+2n)}\Big(\frac{12n^2(2+n)^3}{t^2(\frac{n}{t}+\alpha)^2}-\frac{12n(n+2)^3}{t(\frac{n}{t}+\alpha)}\Big)+\frac{(n+2)^2}{1+2n}\cr&& \Big(-2-2n+\frac{1+2n}{(2+n)^2}+\frac{6n(2+n)}{t^2(\frac{n}{t}+\alpha)^2}\Big).
\end{eqnarray}
The stability of the model during the period of the evolution of the Universe is demonstrated by the observation that the squared sound speed $v_{s}^2$ is always positive in Fig. 10.\\

\section{ Om(z) diagnostic of the proposed model}\label{sec5}
In this paper, we suggest a unique approach $Om(z)$ diagnostic, that allows us to distinguish $\Lambda$CDM from other DE models without using the cosmic EOS directly. This $Om(z)$ diagnostic is a function of red-shift and is built using the Hubble parameter. Due to the change in its slope, this diagnostic is required to refine various dark energy models from the $\Lambda$CDM. The $Om(z)$ is explained as follows for a flat Universe:

\begin{equation}\label{50}
    Om(z)=\frac{\Big(\frac{H(z)}{H_o}\Big)^2-1}{(1+z)^3-1}.
\end{equation}
 Dark energy's behavior can be divided into three categories: quintessence type, which exhibits negative curvature; phantom type, which exhibits positive curvature; and $\Lambda$CDM, which exhibits zero curvature. Fig. 11 shows how $Om(z)$ diagnostic behavior has a positive slope then becomes constant and, as a result, acts for a phantom dark energy model at initial epoch and approaches to dark energy model at present epoch. As a result, the value of $Om(z)$ indicates how the parametrization represents the dark energy model at work \cite{38}.
\begin{figure}
\includegraphics[height=1.5in]{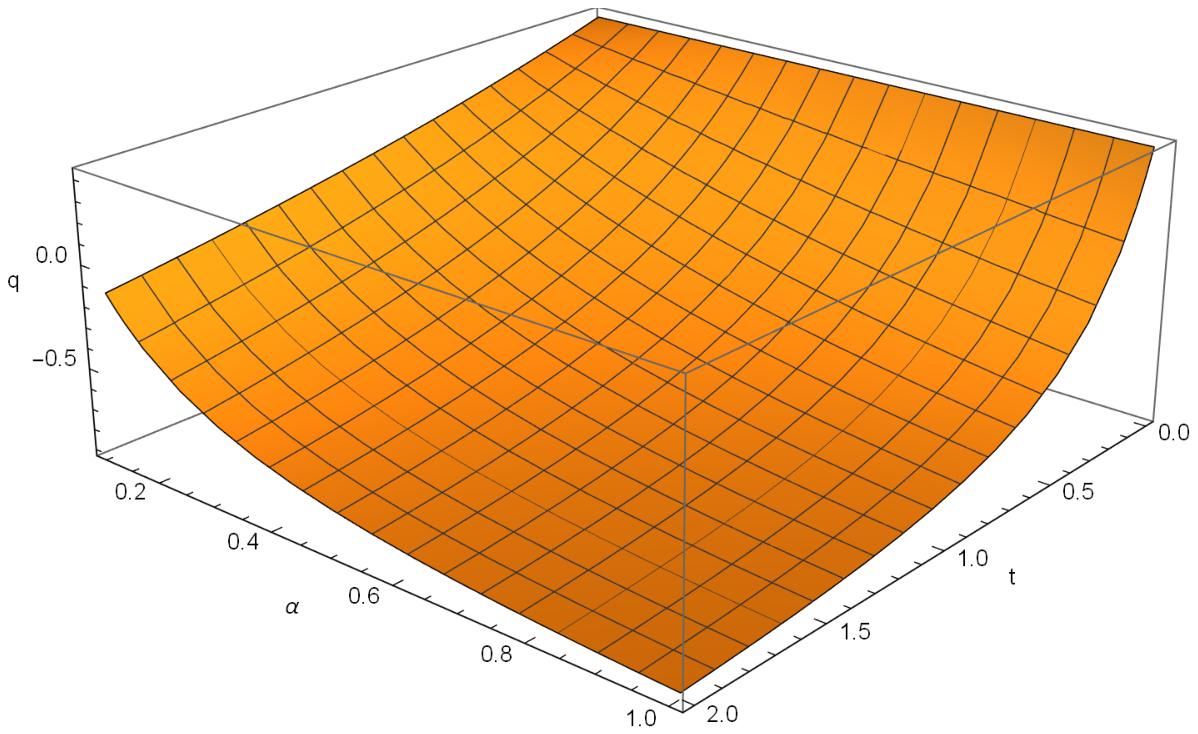}~~~~~
\includegraphics[height=1.5in]{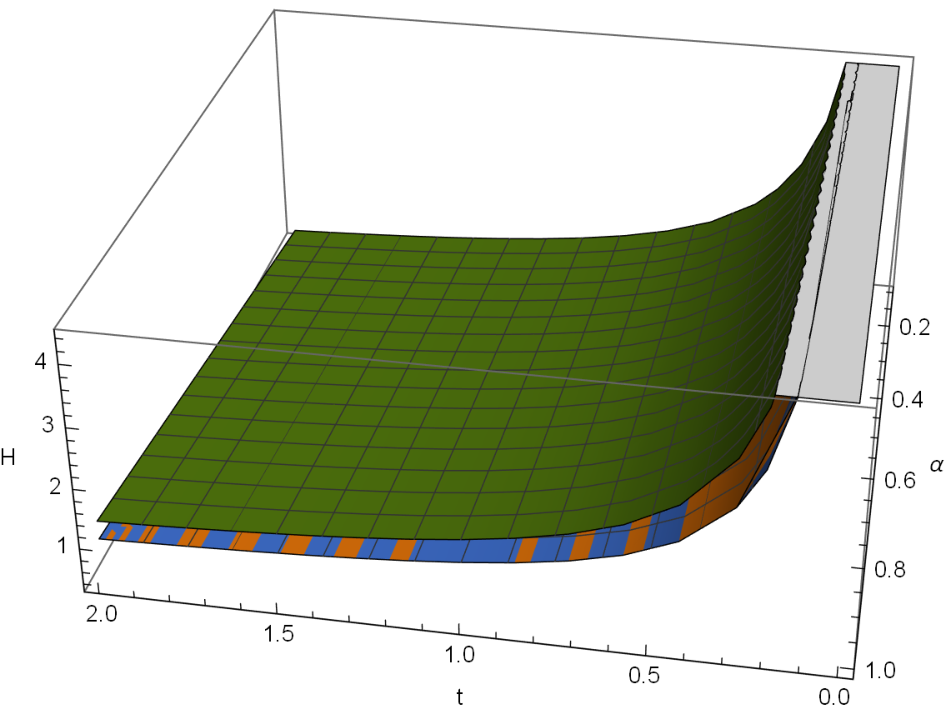}~~~~

Fig.1: Plot of Deceleration parameter vs. time for n=0.5 and $\alpha=0.7$ ~~~Fig.2: Plot of Hubble parameter vs. time for n=0.7 and $\alpha=0.5$

      \end{figure}

       \begin{figure}
\includegraphics[height=1.5in]{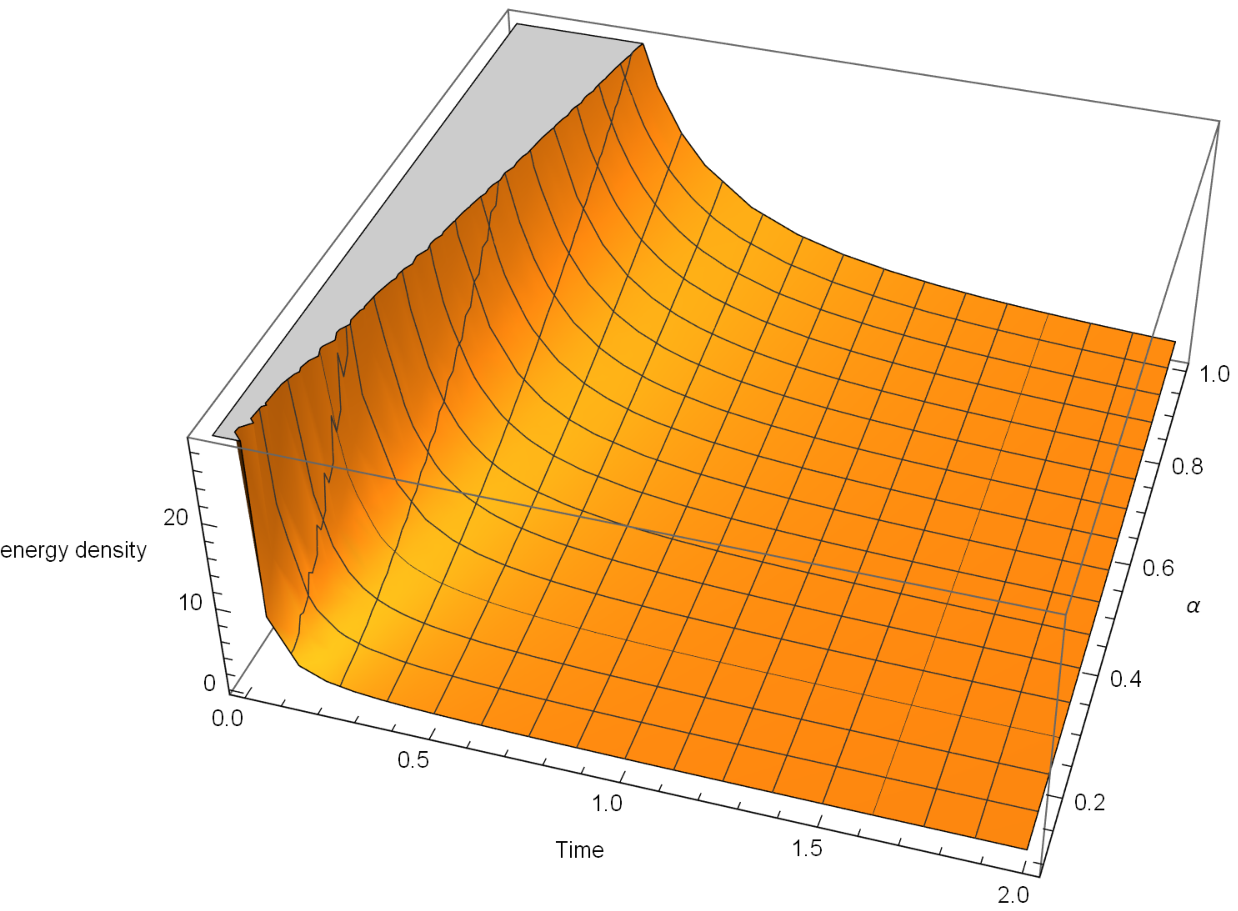}~~~~~
\includegraphics[height=1.5in]{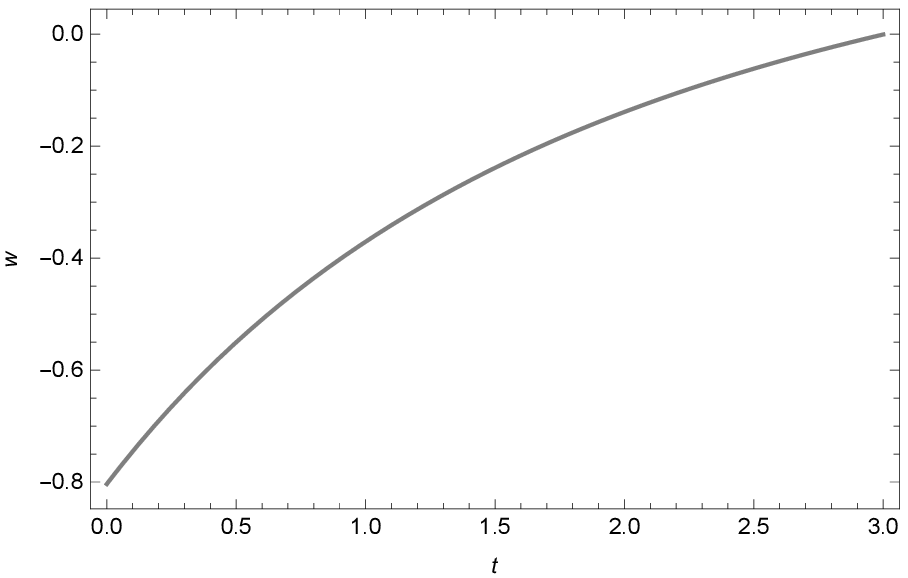}~~~~

Fig.3: Evolution of energy density with time for n=0.7 and $\alpha=0.5$.~~~~~Fig.4: Behaviour of EoS with time for n=0.5 and $\alpha=0.8$.
      \end{figure}

 \begin{figure}
\includegraphics[height=1.5in]{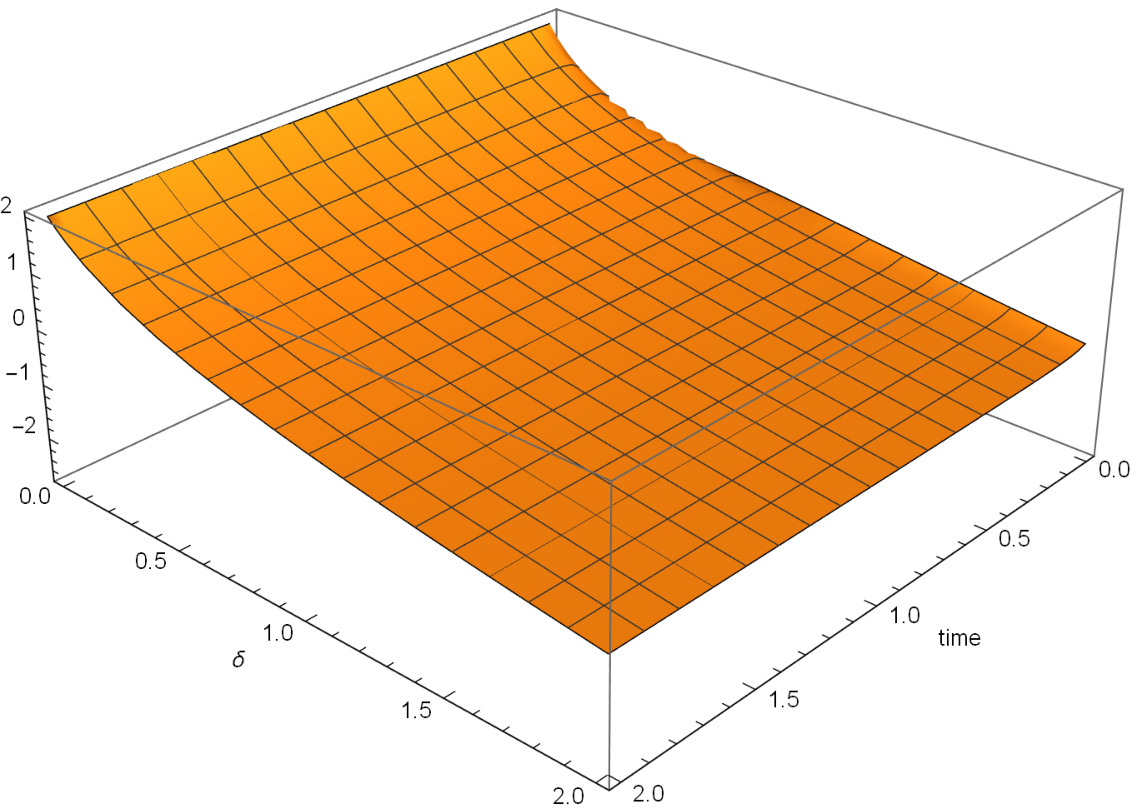}~~~~~
\includegraphics[height=1.5in]{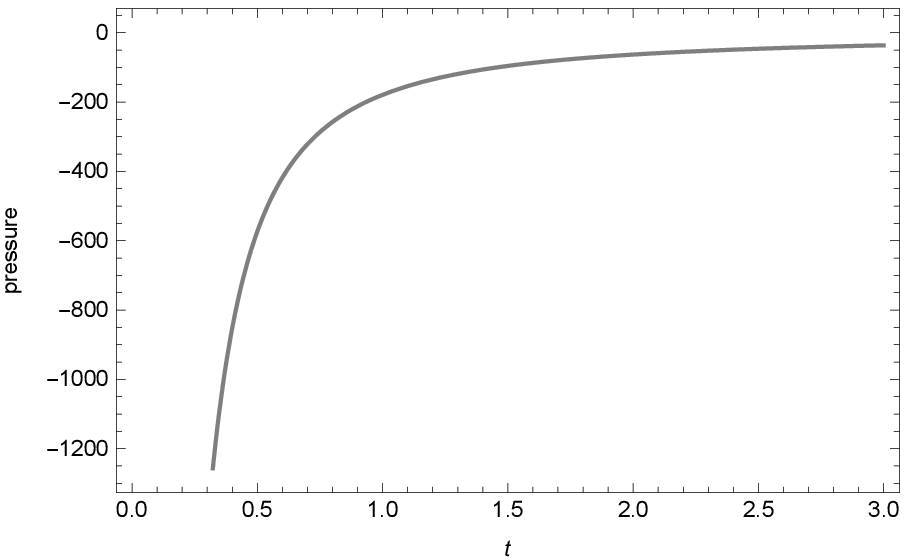}~~~~~

 Fig.5: The behaviour of skewness parameter with time for n=0.1 and $\alpha=0.5$.~~~Fig.6: Evolution of pressure with time for n=0.7 and $\alpha=0.5$.
      \end{figure}

 \begin{figure}
\includegraphics[height=1.5in]{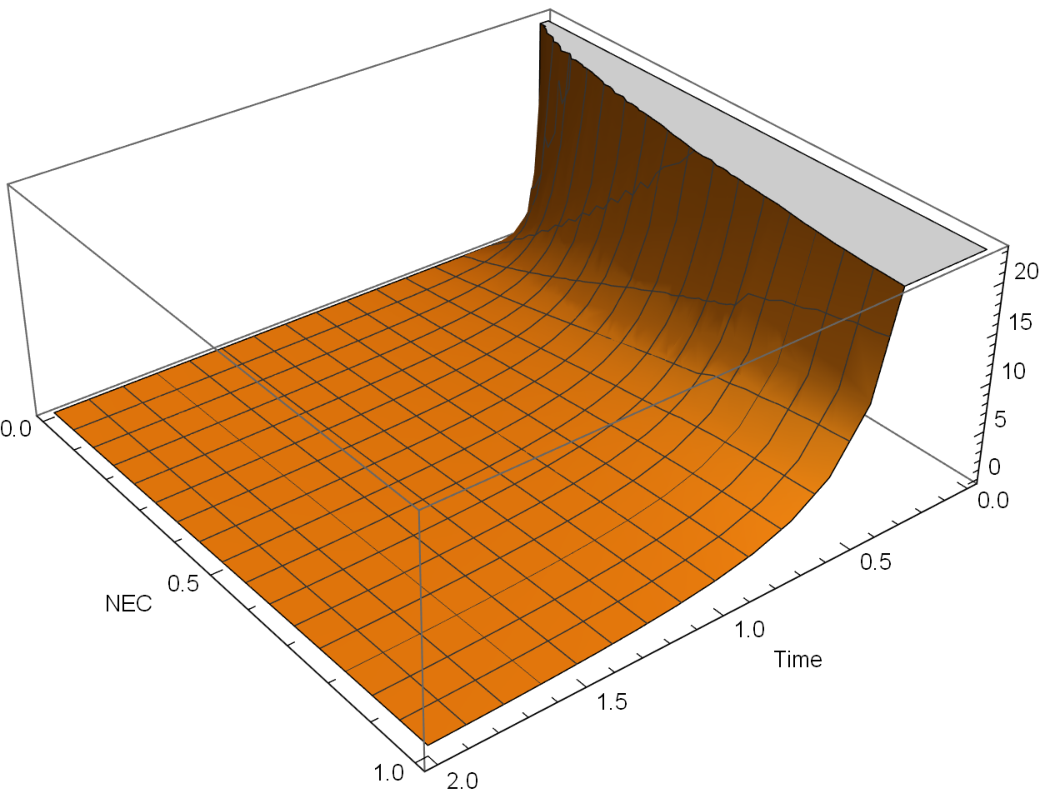}~~~~~
\includegraphics[height=1.5in]{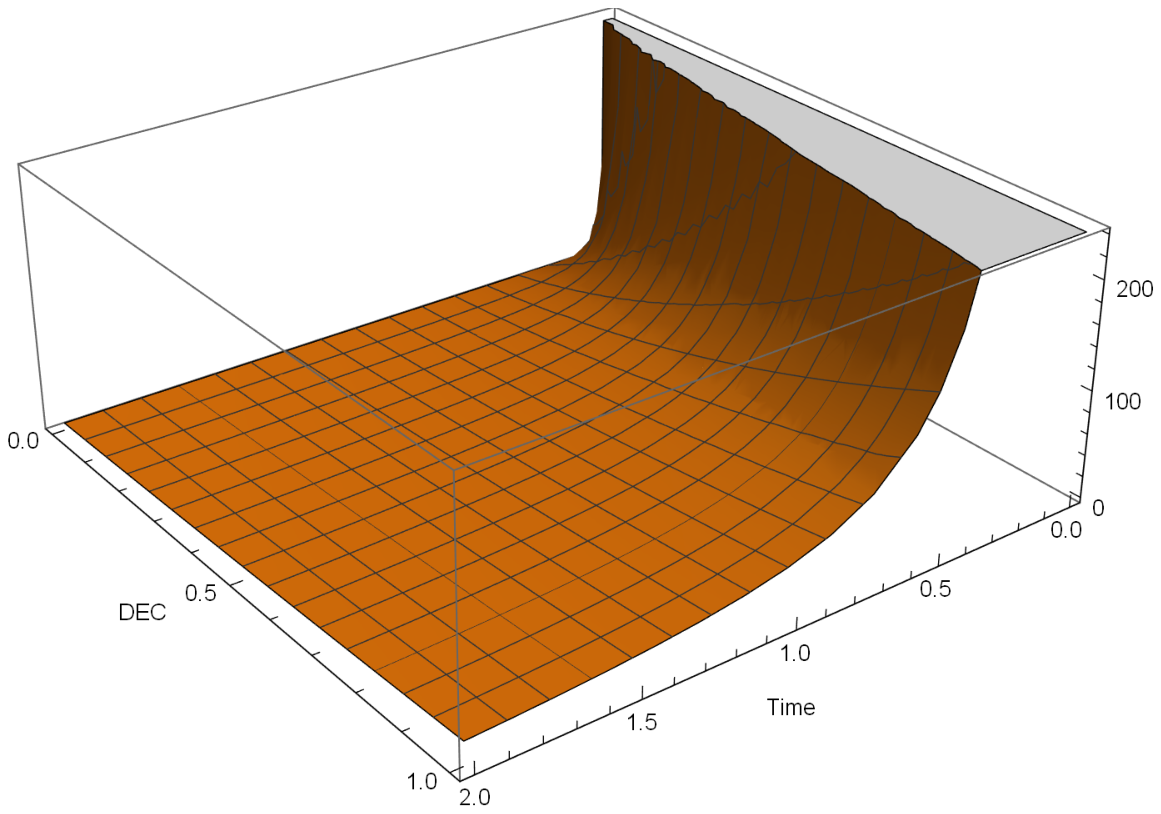}~~~~~

Fig.7: The behaviour of NEC against time for n=0.7 and $\alpha=0.5$. ~~~~Fig.8: The behaviour of DEC against time for n=0.7 and $\alpha=0.5$.

      \end{figure}

\begin{figure}
\includegraphics[height=1.5in]{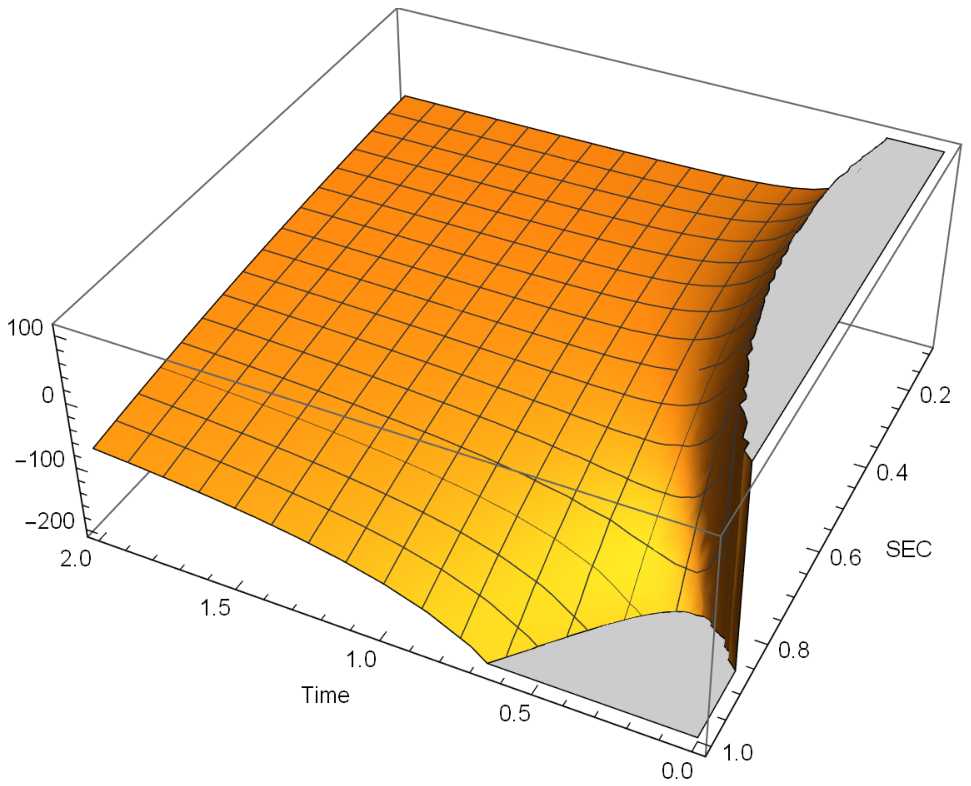}~~~~~
\includegraphics[height=1.5in]{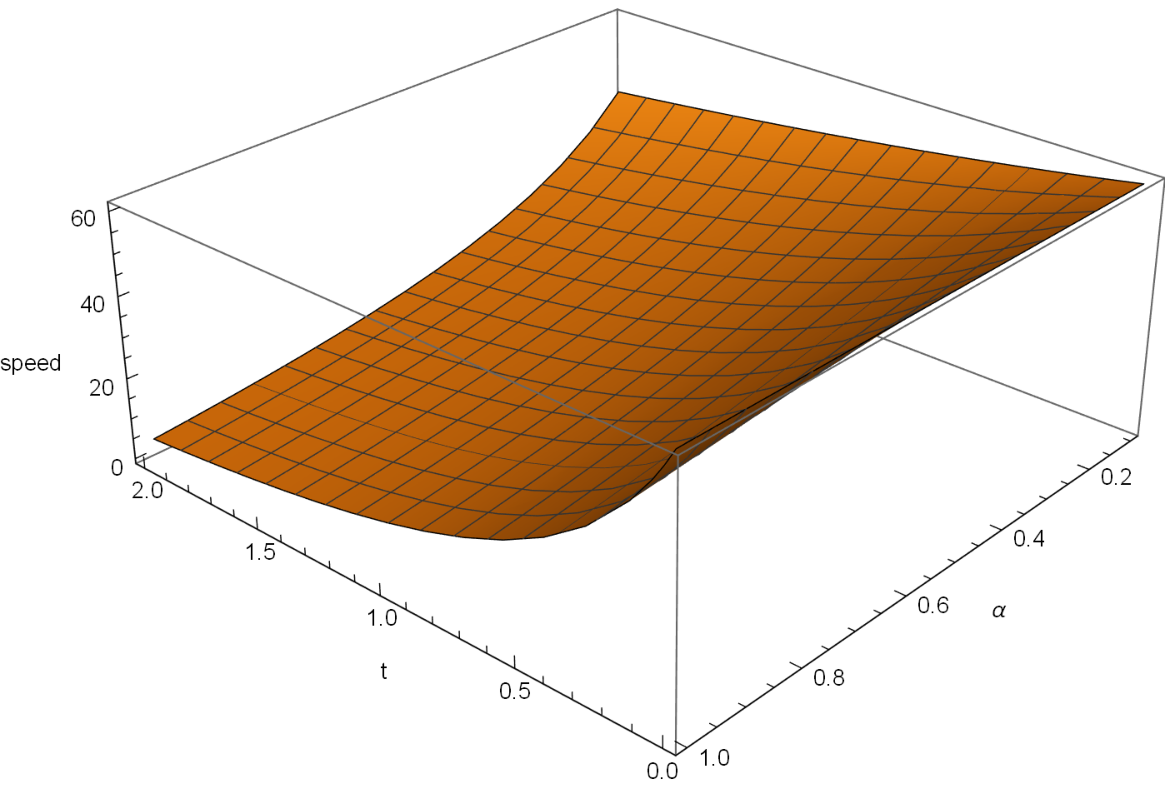}~~~~~

Fig.9: The behaviour of SEC against time for n=0.7 and $\alpha=0.5$.~~~~ Fig.10: The behaviour of speeds of sounds $v_{s}^{2}$ with time for n=0.7 and $\alpha=0.5$.

\end{figure}

\begin{figure}
\includegraphics[height=1.5in]{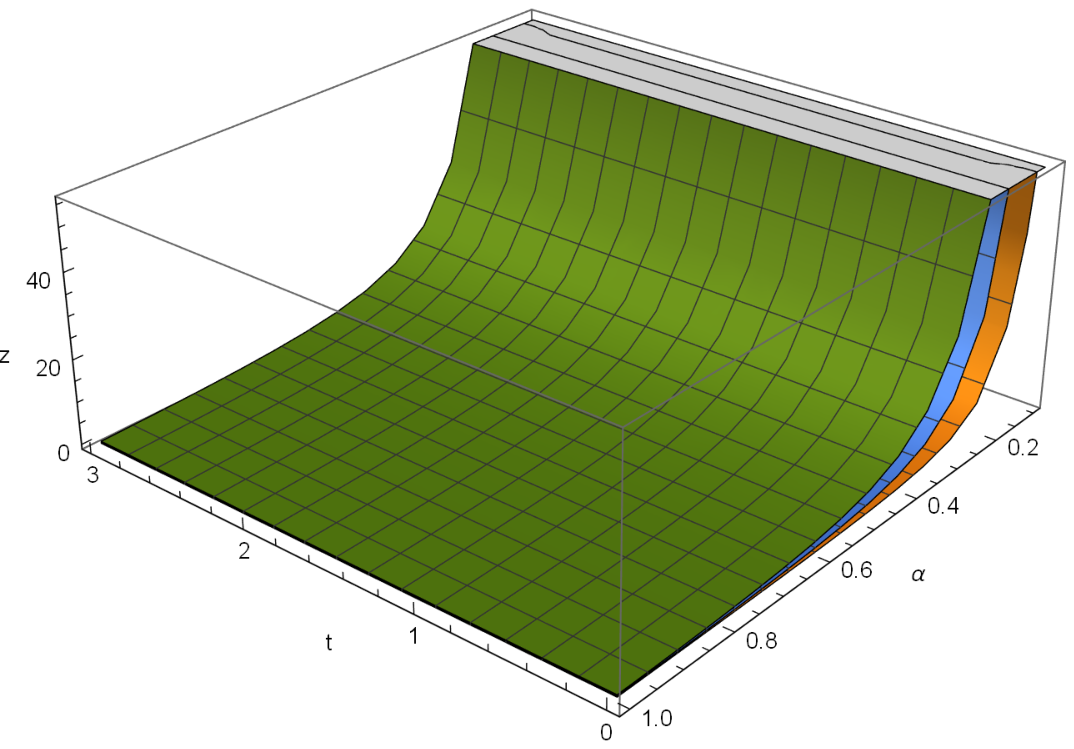}~~~~~

Fig.11: The behaviour of $Om(z)$ diagnostic against time for n=0.7 and $\alpha=0.5$.

\end{figure}

\section{Concluding Remarks}\label{sec6}
The Bianchi type-I Universe was investigated in this study using the well-known modified $f(Q)$ gravity and the presence of the HEL, where the gravitational interaction is visible without the presence of curvature and torsion. The equations of motions for $f(Q)$ gravity in anisotropic homogeneous LRS Bianchi type-I space-time have been considered in this paper. We have taken into consideration $f(Q)=Q$ in order to achieve the results for anisotropic dark energy as an analogy to the GR. Additionally, we have used the HEL to examine the energy density $\rho$, EoS parameter $w$, pressure $p$, and skewness parameter $\delta$. Our research has found that the Universe exhibits a phase of acceleration after a period of deceleration, which is an essential component of the Universe dynamic evolution. The constraint on the anisotropy term demonstrates the Universe late-time acceleration towards isotropy. The EoS parameter in the model is constantly negative as $t \rightarrow \infty$. With the progression of the Universe ages, the energy density $\rho$ is decreasing, and it displays a positive condition that is consistent with observation. The behaviour of skewness parameter which is an useful tool  for checking whether the model is anisotropic or not, in the case of an isotropic Universe, $\delta=0$, and we have found that it tends towards to zero in the future, which confirms that our model will be under isotropic behaviour at late epoch of the Universe. Our model is seen to have an accelerating, non-rotating expansion and lack an initial singularity. The model under consideration satisfies WEC and DEC, but intermittently violates SEC as the Universe expands. Therefore, one can deduce from the energy condition results that SEC violation may lead the Universe to accelerate. In every instance, the square sound speed complies with the constraints $v_{s}^2>1$. As a result, the model under consideration is stable and practical and it starts from phantom model to dark energy model at present epoch. Combining all the findings, it is assumed that the described HEL is physically feasible within the context of modified $f(Q)$ gravity.  Thus, despite its clarity, our manuscript can help readers for better understanding in the study of the dynamics of the cosmos at the beginning of time.

\section*{Declaration of conflict of interest} The author declares no potential conflict of interest or competing interest.

\end{document}